\documentclass{article}

\usepackage{arxiv}
\usepackage{multirow}
\usepackage[utf8]{inputenc} 
\usepackage[T1]{fontenc}    
\usepackage{hyperref}       
\usepackage{url}            
\usepackage{booktabs}       
\usepackage{amsfonts}       
\usepackage{nicefrac}       
\usepackage{microtype}      
\usepackage{lipsum}
\usepackage{graphicx}
 \usepackage{amsmath}
 \usepackage{bbm}
 \usepackage{array}
\graphicspath{ {./images/} }

\title{ Dynamic Treatment Effect Phenotyping through Functional Survival Analysis}

\author{Caterina Gregorio\textsuperscript{1,2}, Giovanni Baj \textsuperscript{3}, Giulia Barbati \textsuperscript{2}, Francesca Ieva \textsuperscript{1,4}\\
\textsuperscript{1} MOX - Modelling and Scientific Computing, Department of Mathematics Politecnico di Milan,\\
Piazza Leonardo Da Vinci 32, Milan 20133, Italy,\\
\textsuperscript{2} Biostatistics Unit, Department of Medical Sciences, University of Trieste, Via Valerio 4$\backslash$1, Trieste 34100, Italy,\\
Piazza Leonardo Da Vinci 32, Milan 20123, Italy,\\
\textsuperscript{3} Department of Mathematics and Geosciences, University of Trieste, Via Valerio 6, 34100, Trieste, Italy,\\
\textsuperscript{4} HDS, Health Data Science center, Human Technopole,  Viale Rita Levi Montalcini 1,  Milan 20157, Italy.}

\begin{document}
\maketitle
\begin{abstract}

In recent years, research interest in personalised treatments has been growing. However, treatment effect heterogeneity and possibly time-varying treatment effects are still often overlooked in clinical studies. Statistical tools are needed for the identification of treatment response patterns, taking into account that treatment response is not constant over time. 
We aim to provide an innovative method to obtain dynamic treatment effect phenotypes on a time-to-event outcome, conditioned on a set of relevant effect modifiers. The proposed method does not require the assumption of proportional hazards for the treatment effect, which is rarely realistic. We propose a spline-based survival neural network, inspired by the Royston-Parmar survival model, to estimate time-varying conditional treatment effects. We then exploit the functional nature of the resulting estimates to apply a functional clustering of the treatment effect curves in order to identify different patterns of treatment effects.  The application that motivated this work is the discontinuation of treatment with Mineralocorticoid receptor Antagonists (MRAs) in patients with heart failure, where there is no clear evidence as to which patients it is the safest choice to discontinue treatment and, conversely, when it leads to a higher risk of adverse events. The data come from an electronic health record database. A simulation study was performed to assess the performance of the spline-based neural network and the stability of the treatment response phenotyping procedure. In light of the results, the suggested approach has the potential to support personalized medical choices by assessing unique treatment responses in various medical contexts over a period of time.

\end{abstract}


\section{Introduction}

\begin{figure}
    \centering
    \includegraphics[width=13cm]{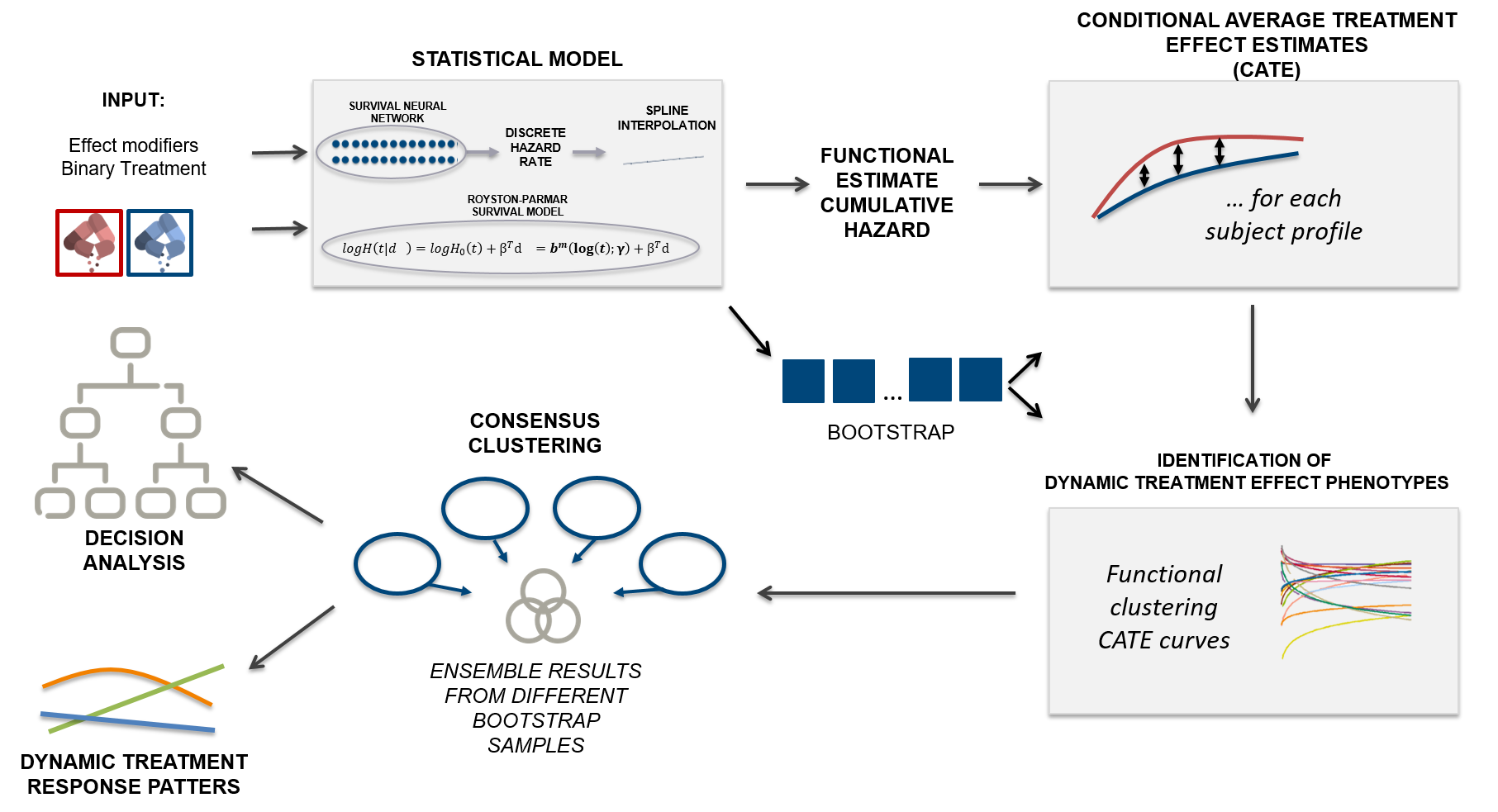}
    \caption{Summary of the proposed method to obtain dynamic treatment effect phenotypes.}
    \label{fig:abstract}
\end{figure}

The field of personalized treatments has been experiencing a noticeable surge in research interest. This can be attributed to the growing recognition that the one-size-fits-all approach has significant limitations. Nonetheless, despite this heightened focus, there remains a tendency to neglect the potential variations in treatment effects over time when considering time-to-event outcomes. When studying the problem of obtaining personalized treatment decisions in order to minimize the subject's risk of adverse events, it is important to acknowledge how time affects the response to the treatment. In fact, the treatment response that we observe is typically the result of multiple biological mechanisms that a treatment triggers in an individual. These may differ in their timing, magnitude and nature, as some may be beneficial and others may be side effects. Whereas there is a growing literature on optimal treatment rules using machine-learning algorithms such as reinforcement learning \cite{Sutton2009ReinforcementIntroduction}, the aim of this work is to provide a method to identify common treatment effect patterns over time to characterize the dynamic response of individuals to treatments. The rationale is that such a method can 1) provide valuable information on the whole time horizon of interest without restricting the attention to a single point in time and, 2) inform on the timing and direction of the treatment response of individuals. As a result, the medical expert can make a judgment of the best treatment decision to take for a specific patient profile. For example, this approach can help distinguish subject profiles that have an early vs. late response to a treatment, or it can provide evidence on whether the effect of a treatment or exposure tends to fade away or increase over time. According to the specific medical context and specific decision problem, this information can have different implications, and it is important for the decision maker to have them before choosing one treatment alternative over the other. 
In order to do so, the Average Treatment Effect conditioned on the effect modifiers, the so-called Conditional Average Treatment Effect (CATE), is estimated taking into account its possibly time-varying nature.  To characterize the different treatment-effect phenotypes over time, a clustering procedure is performed considering the time dimension of the CATE. Functional Data Analysis (FDA)\cite{Ramsay2005} is a part of statistics that collects methods to model data that can be thought of as being realizations of functional random variables. Usually, starting from discrete observations, the first step consists of obtaining functional estimates for each unit using smoothing. 
In survival analysis, the Royston-Parmar survival model\cite{Royston2002FlexibleEffects} allows retrieving proper functional estimates of the survival curves by modelling the baseline hazard with natural cubic splines. The spline-based model allows obtaining smooth estimates for the cumulative hazard and survival functions, meaning that derivatives, i.e. hazard, are also available. Similarly to some FDA methods, smoothing is achieved with basis expansion, which allows controlling for the degree of smoothing without imposing strict functional forms. Moreover, the model can also be extended to include time-dependent effects. The main limitation of this method is that model selection is challenging when possible interactions among covariates and time-dependent effects are present. As an alternative, we propose to use a survival neural network approach to estimate the CATEs enabling us to easily capture all relevant interactions and time-dependent effects. Among the different methods, we consider the Logistic-Hazard method, also called Nnet-survival\cite{Gensheimer2019ANetworks.}.  It is a discrete-time method that models the hazard function non-parametrically. Inspired by the Royston-Palmar model and FDA, we propose to interpolate the (discrete) neural network's predictions with natural cubic splines, to obtain smooth estimates of CATEs. \\
In FDA, a natural way to group and classify curves in an unsupervised manner is functional clustering. Here, we used it to identify relevant phenotypes that describe different responses to treatment over time, which can then be mapped to the different subjects' profiles.\\
The application that motivated this work is the discontinuation of treatment with Mineralocorticoid receptor Antagonists (MRAs) in patients with heart failure, where there is no clear evidence as to which patients it is the safest choice to discontinue treatment and, conversely, when it leads to a higher risk of adverse events. The data come from an Italian Electronic Health Record Database. The aim of the study is to provide evidence for the development of individualised recommendations for the discontinuation of MRAs according to the relevant clinical characteristics of the subjects.
In Section 2 we describe the statistical methods. Section 2.1 introduces the notation and the theoretical framework. Details of how CATE can be estimated through either the spline-based model or the spline-based survival neural network are given in Section 2.2. The clustering procedure and the identification of the different phenotypes are reported in Section 2.3. Section 3 is devoted to the simulation study, whereas the description of the data used, the motivating problem and the results of the application are reported in Section 4. Finally, discussion and conclusions are reported in Section 5. A summary of the different
steps of the method is reported in Figure \ref{fig:abstract}.

\section{Methods}


\subsection{Notation and Formal Framework}
We let $D$ and $C$ the event of interest and censoring time respectively. We assume that $C$ is non-informative with respect to the event time. For each subject, we only observe the couple $min(D,C)$, $\delta=\mathbbm{1}(D\le C)$.
The binary random variable indicating the covariate of interest, i.e., the treatment or exposure, is denoted by $Z$. Let $\textbf{X}$ be a  $p$-dimensional vector of observable covariates with $p>1$.

We define as $D^{(1)}$ the potential outcome if the treatment/exposure is received and $D^{(0)}$ the corresponding potential outcome without treatment/exposure.

We are interested in estimating the Conditional Average Treatment Effect (CATE) at each time $t \in [0,w]$ on a subset of the covariate vector  $\textbf{X}$, denoted as $\textbf{X}_1$, that contains the treatment-effect modifiers :

\begin{equation}
    \tau_{\textbf{x}_1}(t)^{[0,w]}=E\{f[g(D^{(1)};t),g(D^{(0)};t)]|\textbf{X}_{1}=\textbf{x}_{1}\}
    \label{CATE}
\end{equation}

where $g(\cdot)$ can be either the survival, hazard or cumulative hazard function and $f(\cdot)$ is a measure of effect, typically either the ratio or the difference. In the following, simplifying the notation, we omit the reference to the specific time horizon of interest, denoted as $[0,w]$, in $\tau_{\textbf{x}_1^{[0,w]}}(t)$.

To identify the CATE, we need to assume the treatment groups are \textit{conditionally exchangeable}:

\begin{equation}
    [D^{(1)},D^{(0)}]\perp Z|\textbf{X}_{1}
\end{equation}

This assumption is also known as ‘‘ignorable treatment assignment''.  If the observed covariates contained in $\textbf{X}$ but not in $\textbf{X}_1$ act as confounders,  then:

\begin{equation}
    [D^{(1)},D^{(0)}]\perp Z|\textbf{X}
\end{equation}

In the latter case, to ensure conditional exchangeability,  a propensity score method such as matching or Inverse Probability of Treatment Weighting needs to be applied.

The steps of the method to identify treatment-effect phenotypes consist of 1) estimating the CATE as in equation \ref{CATE},
2) identifying the grouping structure of the CATE curves through an unsupervised clustering method  3) characterizing the different behaviours in terms of response to the treatment 4) mapping the different subject profiles to their corresponding treatment-response behaviour.

\subsection{Estimation of functional CATEs}

In the Royston-Parmar Survival  Model (R-PSM), the logarithm of the baseline cumulative hazard function is modelled as a natural cubic spline function of log time:

\begin{equation}
    log H(t|\textbf{d})= log H_0(t)+\bbeta^T \textbf{d}=\textbf{b}^m(log(t);\bgamma)+\bbeta^T \textbf{d}
    \label{RPm}
\end{equation}

where $\textbf{b}^m(t;\bgamma)$ is a natural cubic spline with $m$ knots and $\textbf{d}^T=[ z \; \textbf{x}_1 \; z\times \textbf{x}_1]^T$. The flexibility in the form of the baseline hazard is given by the number of internal knots. The inclusion of interactions between covariates and the treatment variables enables the presence of heterogeneity in the treatment effects.
This model can also be easily extended to include time-varying coefficients by modelling the spline coefficients in the function of the covariates for which we want a time-varying effect. The general model with time-dependent effect can be written as: 

\begin{equation}
    log H(t|d)= \gamma_0 + \bgamma \textbf{b}^m(log(t))+\bbeta^T \textbf{d}
\end{equation}

where $dim(\bgamma)=m+1$ and the $jth$ component of $\bgamma$ is:

\begin{equation}
\gamma_{j}=
    \begin{cases}
    \gamma_{j0} & k=0 \text{ Proportional Hazard model} \\
    \gamma_{j0}+\sum_{r=1}^k\gamma_{jr}d_r  & k\ge 1 \text{ Non-Proportional Hazard model for } d_1,\cdots d_r \\
    \end{cases}
\end{equation}

The parameters in the model can be estimated using Maximum Likelihood (ML) and their uncertainty can be evaluated using standard ML asymptotic theory. The degree of freedom of the splines is typically chosen with the AIC or BIC and through visual comparison of the fitted hazard vs. non-parametric estimates. This model can be fitted in \texttt{R} using the package \texttt{flexsurv} \cite{Jackson2016Flexsurv:R.} or \texttt{stpm2} \cite{X.-R.Liu2018ParametricModels}.

A model-free approach consists of using a survival neural network that estimates the hazard function. Specifically, we consider the  Nnet-Survival method, also called Logistic-Hazard, which parametrizes the discrete-time hazard rate with a neural network and optimizes a survival likelihood expressed in terms of the discrete hazards non-parametrically \cite{Gensheimer2019ANetworks.}. In the Nnet-Survival method, follow-up time is divided into $h$ intervals which are left-closed and right-open. The contribution of a generic time interval $j$ to the overall log-likelihood is:

\begin{equation}
    \sum_{i=1}^{d_j}=log(h^i_j)+\sum_{i=d_j+1}^{r_j}log(1-h^i_j)
    \label{lliknet}
\end{equation}

where $r_j$ is the number of subjects at risk before the beginning of the interval, $d_j$ is the number of subjects experiencing the event during the interval and, $h^i_j$ is the hazard for an individual $i$ during the time-interval $j$.
The loss function in Equation \ref{lliknet} comes from classic discrete-time survival models, and its use in a neural network context is well justified by survival analysis theory.
Furthermore, it naturally incorporates a time-varying baseline hazard rate and time-varying effect, since each time interval output node is fully connected to the last hidden layer’s neurons. As architecture, we used a fully connected network with 2 hidden layers of 32 units each, and ReLU as a non-linear activation function.
The neural network gives, for each subject profile, a $h$-dimensional output corresponding to a discrete set of hazard rates, one for each interval. We then propose to interpolate the hazard curves using natural cubic splines. In this way, similar to the previous model we obtain a smooth estimate of $H(t|d)$ that depends in this case on a non-linear function of the covariates $x_1$ and the treatment indicator $z$. This neural network was implemented in \texttt{python} using \texttt{pycox} \cite{Kvamme2021ContinuousNetworks} and \texttt{pyTorch} \cite{paszke2017automatic}.

From both the Royston-Parmar Survival Model (R-PSM) and the Spline Nnet-Survival (SNnet-S), we obtain a functional estimate of the CATE as defined in equation \ref{CATE}, by predicting the measure of effect of interest, $g(\cdot)$, for each combination of the covariates $x_1$ and by comparing it between the two treatment strategies $z=0$ and $z=1$ using the chosen function $f(\cdot)$. The estimation of the CATEs is only the first step of the procedure that leads to the identification of the treatment effect phenotypes. To address the inherent uncertainty in these estimates in subsequent steps and enhance the overall reliability of our findings, we employ resampling techniques to quantify the variability in CATE estimates. For the model-based method, a simulation approach based on parametric bootstrap \cite{Mandel2013Simulation-BasedDerivatives} is applied. For each bootstrap sample, $b=1,\cdots,B$ the methods consist of:

\begin{enumerate}
    \item sampling $\bgamma^{(b)}$ from the asymptotic Normal distribution of the Maximum Likelihood of the regression parameters,
    \item obtaining the corresponding estimates of CATEs $\hat{\tau}_{\textbf{x}_1}^{(b)}(t)$
\end{enumerate}

On the other hand, for the neural network, non-parametric bootstrap is used. Consequently, for every combination of the covariates $\textbf{x}_1$ under investigation, we generate a distribution comprising $B$ estimates of CATEs that we will employ in the following step for the identification of the treatment effect phenotypes.

\subsection{Identification of treatment effect phenotypes over time}

Functional clustering is used to find a grouping structure for  $\hat{\tau}_{\textbf{x}_1}^{(b)}(t)^{[0,w]}$, $b=1,\cdot,.,B$ for each combination of the covariates $\textbf{x}_1$ of interest in order to aggregate profiles of individuals in clusters with a common response to the treatment exposure over time. The functional clustering procedure is repeated on each bootstrap sample to take into account the uncertainty in the estimation of the CATEs and improve the stability of the procedure. In FDA, similarly to multivariate data analysis common clustering methods consist of k-means clustering, hierarchical agglomerative clustering or density-based clustering such as the DBSCAN. These, have all been extended to take into account that observations of the group are functional objects \cite{Tarpey2003ClusteringData}. The implementation used in the \texttt{R} package \texttt{fdacluster} \cite{Stamm2023Fdacluster:Data} has been considered in this work. Specifically, we used functional k-means clustering, choosing the L2 distance between the CATE estimates of the different subject profiles:

\begin{equation}
    d_2\{\hat{\tau}_{l}^{(b)}(w),\hat{\tau}_{m}^{(b)}(w)\}=\sqrt{\int_0^w\{\hat{\tau}_{l}^{(b)}(s)-\hat{\tau}_{m}^{(b)}(s)\}^2 ds}
\end{equation}

for $b=1,\cdots,...,B$ and each couple of subject profiles $l,m$.
To aggregate the clustering results obtained on the different bootstrap samples, either a simple majority vote or an ensemble method based on consensus clustering can be used.  Given a measure of similarity  (or agreement) between the results of two clustering, consensus clustering aims at maximizing the average similarity. It is possible to distinguish between medoid consensus clustering \cite{Strehl2003RelationshipBasedCA}  and soft consensus clustering \cite{Gordon2001FuzzyPartitions}. According to medoid consensus, final clustering is sought over the set of the base clusterings. In soft consensus clustering, it is possible to assign objects to several groups with varying degrees of “membership”.  These clustering ensemble methods are all implemented in  \texttt{R} package \texttt{clue}\cite{Hornik2005AEnsembles}.  
K-means clustering requires choosing the number of clusters. The silhouette values are typically employed for this task. Alternatively, the re-sampling procedure through bootstrap allows us to consider the mean internal agreement between the results of the clustering on the different bootstrap samples to assess the stability of the results with different numbers of clusters. Finally, clusters can be interpreted through the visual inspection of the clustering consensus centroids. These were obtained as the functional median obtained using the Modified Band Depth \cite{Lopez-Pintado2009OnData,RJ-2019-032} of the CATEs contained in each final consensus cluster.

\section{Simulation Study}
\subsection{Design}

We performed a simulation study to evaluate the performance of the proposed method. In the first part of the simulation, we compare the performance of the SNnet-S with the R-PSM.  Secondly, we want to assess the stability of the identification of treatment effect patterns through functional clustering.  We considered a vector of 5 binary effect modifiers, and we specified a data-generating model based on the spline-based parametric survival model with time-varying effects and several interactions among the covariates. According to this model, the baseline hazard was specified considering a spline with 1 knot placed at $t=10$. 100 datasets were simulated under four different sample size scenarios (n=1 500, 5 000, 10 000, 50 000). As a measure of effect in the simulation, we considered the ratio between cumulative hazard.

\subsection{Results}

\begin{figure}
    \centering
    \includegraphics[width=17cm]{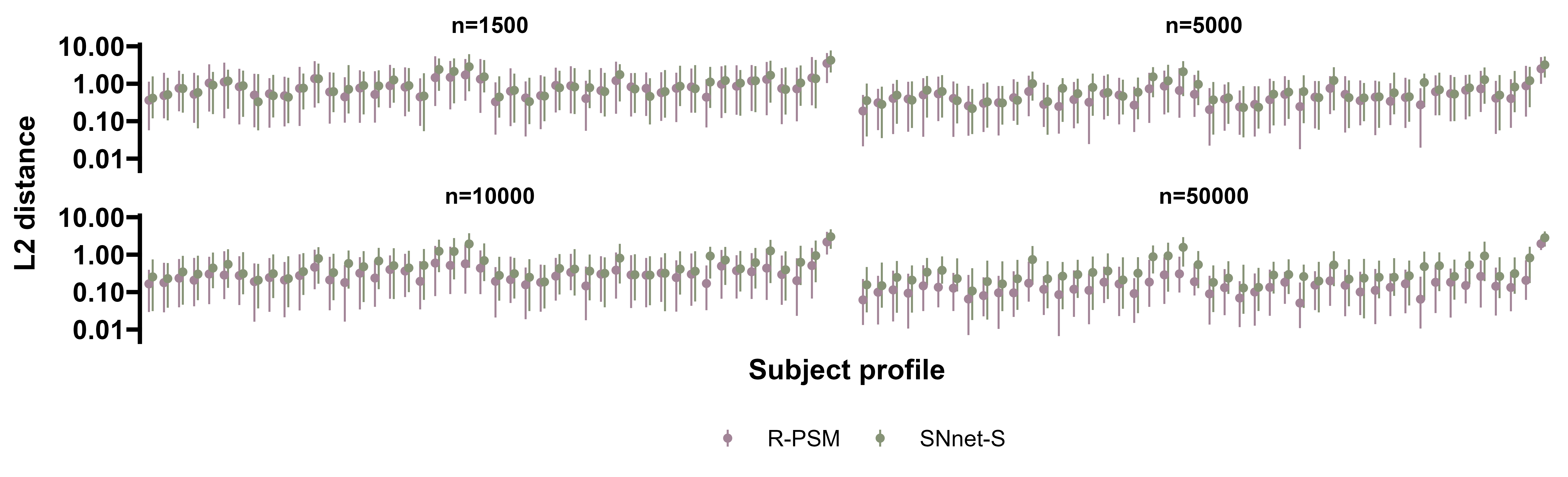}
    \caption{Comparison of the median L2 distance with 95\% CI between the estimated cumulative hazard of a correctly specified R-PSM (pink)  and SNnet-S (green), relative to the true data generating model.}
    \label{fig:sim1}
\end{figure}

The L2 distance between the cumulative hazard of the true data-generating model and the one estimated with either the correctly specified R-PSM or the SNnet-S approach was calculated for each combination of the covariates and binary treatment indicator. The results for the six scenarios are reported in Figure \ref{fig:sim1}. For all sample sizes and all subject profiles, the SNnet-S and the R-PSM reach a similar performance. It is important to note that the performance of R-PSM is conditioned on the fact that we are able to select the correct model.

Established the satisfying performance of the SNnet-S approach, we continued with this approach and performed the functional clustering considering different numbers of clusters $nclust=2,3,4$. 100 was used as the number of bootstrap samples.
As previously reported, to ensemble the clustering results obtained in the different bootstrap samples a naive majority vote and consensus clustering methods were considered. Specifically, here we considered three medoid consensus methods and two soft least square consensus methods that differ for the (di-)similarity measure used: Euclidean, Manhattan and Rand are considered for the medoid consensus and, Euclidean and Manhattan are considered for the least square soft consensus.

\begin{figure}
    \centering
    \includegraphics[width=10cm]{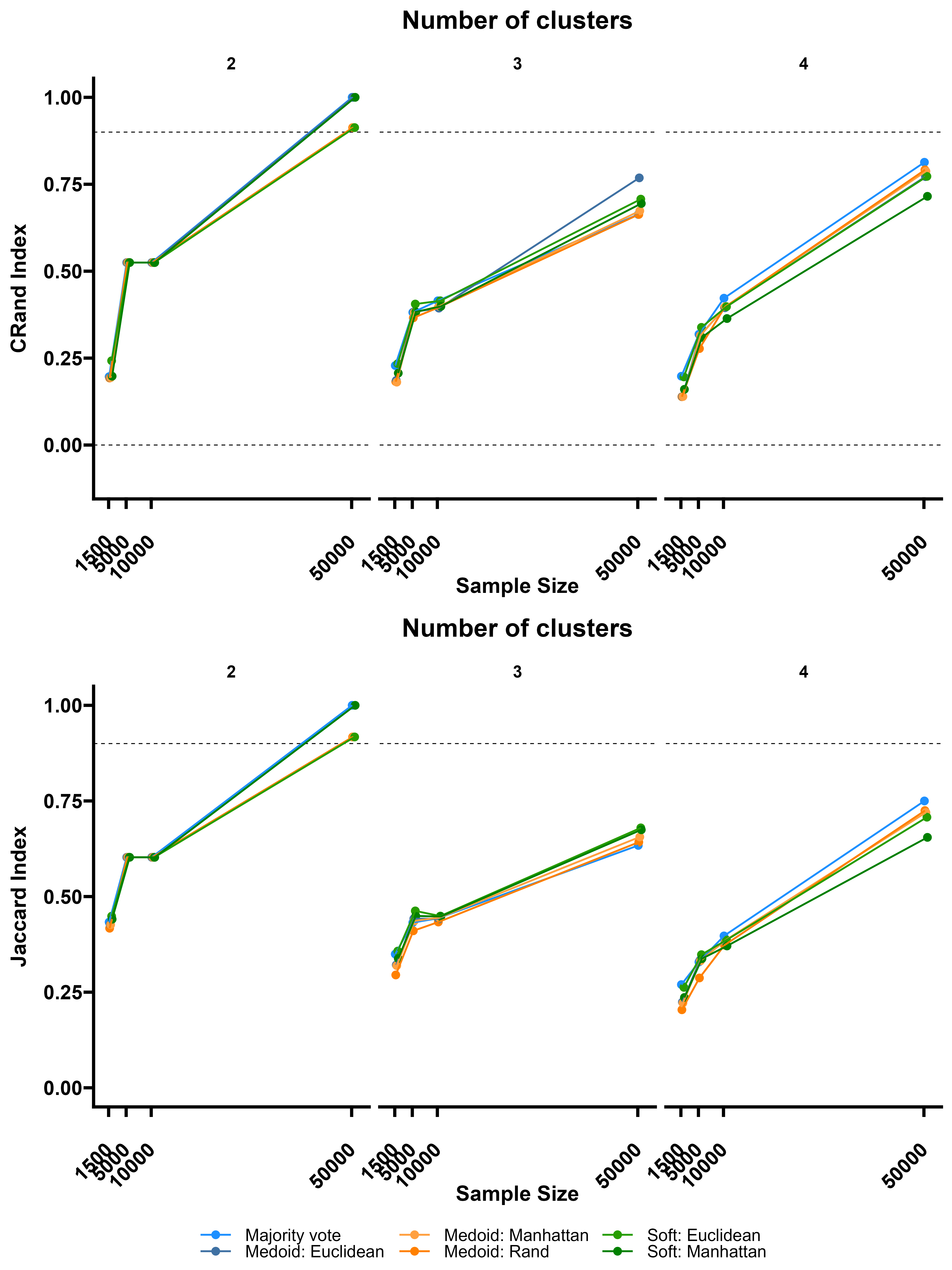}
    \caption{Comparison of the agreement according to the CRand Index (top panel) and the Jaccard Index (bottom panel) according to different bootstrap ensemble methods.  }
    \label{fig:sim2}
\end{figure}

The results of the clustering obtained on cumulative hazard ratio curves of the data-generating model are considered the “gold standard” and they are compared in terms of agreement across the different ensemble bootstrap methods by means of the Jaccard Index and the Corrected Rand (CRand) Index. The Jaccard index compares the ratio of the numbers of distinct pairs of subject profiles in the same class in both partitions and in at least one partition, respectively, and it can take values from 0 to 1. The CRand Index calculate the rate of distinct pairs of subject profiles both in the same class or both in different classes in both partitions, and it is corrected for agreement by chance. The latter index can take values from -1 to 1 and negative values indicate that agreement is less than what is expected from a random result.
In Figure \ref{fig:sim2} we can observe that the agreement increases with the sample size. Importantly, the results are robust to the choice of the clustering consensus method. The agreement is higher for the two clusters, and this is due to the specific data-generating model used. In particular, considering two clusters, the method achieves an agreement with the “true” clustering above 0.90 according to both indices and all clustering consensus methods on the scenario n=50 000. In Table \ref{boot_agree} we report the mean agreement among the clustering results in the 100 bootstrap samples. As expected, a higher internal agreement among clustering samples corresponds to a higher final performance of the method. For two clusters and n=50000, the mean internal agreement is above 0.5.

\begin{table}
\centering
\caption{Concordance of clustering results between bootstrap samples}
\label{boot_agree}
\begin{tabular}{llcc}
\hline
Sample size & Number of clusters & CRand Index & Jaccard\\

\hline
& 2 & 0.06 & 0.38\\
\cline{2-4}
 & 3 & 0.05 & 0.25\\
\cline{2-4}
\multirow{-4}{*}{\raggedleft\arraybackslash 1500} & 4 & 0.04 & 0.19\\
\cline{1-4}
 & 2 & 0.22 & 0.45\\
\cline{2-4}
 & 3 & 0.19 & 0.32\\
\cline{2-4}
\multirow{-3}{*}{\raggedleft\arraybackslash 5000} & 4 & 0.16 & 0.25\\
\cline{1-4}
 & 2 & 0.30 & 0.49\\
\cline{2-4}
 & 3 & 0.25 & 0.36\\
\cline{2-4}
 \multirow{-3}{*}{\raggedleft\arraybackslash 10000}& 4 & 0.21 & 0.28\\
\cline{1-4}
 & 2 & 0.57 & 0.65\\
\cline{2-4}
 & 3 & 0.46 & 0.50\\
\cline{2-4}
\multirow{-3}{*}{\raggedleft\arraybackslash 50000} & 4 & 0.41 & 0.41\\
\bottomrule
\hline
\end{tabular}

    \label{tab:my_label}
\end{table}

\section{Application}
\subsection{Clinical problem \& Data}

Treatment of Heart Failure relies on several life-saving pharmacological therapies. Among them, MRAs are one of the cornerstones of therapy in heart failure, yet is one that is most often discontinued by cardiologists out of fear of adverse events, e.g. alteration of potassium \cite{Maggioni2016TheDatabase,Komajda2016PhysiciansSurvey.}. However, there is no clear evidence regarding which patients' side effects overcome benefits in terms of the risk of hospitalisation or death. As a consequence, treatment decisions are often made on a subjective basis. The aim of this analysis is to provide the following medical decision: for which profiles of patients discontinuing MRAs during the first year after initiation of the therapy is the safest choice and contrarily, for which type of patient discontinuation leads to higher risks in terms of risk of hospitalization and/or death? According to European guidelines, in deciding whether to discontinue therapy with MRAs cardiologists should consider episodes of hyperkalaemia (high potassium) since starting therapy, age, diabetes, New York Heart Association (N.Y.H.A class), renal function and ejection fraction. Therefore, these variables were assumed to be the effect-modifying variables. In addition,  other available variables were considered as possible confounders and were adjusted for in the analysis via propensity score matching.

Data were obtained by the interrogation of the administrative regional health data of Friuli Venezia Giulia Region in the Northern part of Italy, integrated with data derived from the Outpatient and Inpatient Clinic E-chart (Cardionet \textregistered). This integrated database constitutes the Trieste Observatory of Cardiovascular Diseases. Specifically, this was a cohort observational, non-interventional study involving patients living in the Trieste who had a Heart Failure diagnosis between January 2009 and December 2020, had at least one cardiological evaluation, two potassium measurements and, were observed for at least one year after having started therapy with MRAs.  For the identification of HF patients, the following steps will be followed. First, a search in the electronic medical records, using appropriate keywords (Heart Failure, Chronic Heart Failure, Systolic Heart Failure, Diastolic Heart Failure) to select patients with HF-related clinical findings. In order to avoid any diagnostic underestimation, data from the medical E-chart were combined with the discharge codes of any previous hospital access (based on the standard nomenclature of the ICD-9 CM) and/or interventional procedures for HF patients (i.e. ICD implantation). Subsequently, prospective cases were manually reviewed by clinicians, to validate the diagnosis of HF using the criteria established in 2016 by the European Cardiology Society. The cohort was followed from the index date, defined as the first date of purchase of MRAs, until an event of cardiovascular hospitalization, death or the end of the follow-up (administrative study closure date, fixed at 31 December 2020).  The database has been previously described in the literature \cite{Iorio2019}. As previously pointed out, all patients had at least one year of observation. The first year will be used as a pre-follow-up period to define the treatment group: persistently treated with MRAs in the first year of therapy vs. therapy with MRAs discontinued during the first year. As a consequence, the analysis will consider the follow-up as the time since the end of the first year of observation.

\subsection{Study Cohort}

1555 subjects were included in the analysis. The outcome of interest was the first event between an event of hospitalization due to a cardiovascular reason or death.
At 2 years of follow-up, the Kaplan-Meier estimate of the event-free probability was equal to 0.66 (95\% CI: 0.63-0.68). 64\% of the patients are older than 75. 39\% present Heart Failure with reduced left-ventricular ejection fraction (<50\%) and 14\% have severe symptoms (N.Y.H.A III or IV)   Moreover, 31\% have diabetes and 47\% have Chronic Kidney Disease (CKD). During treatment with MRAs, \%11 had hyperkalemic episodes.

\subsection{Defining the treatment variable: discontinuation of therapy with MRAs}

Using drug prescription data coming from administrative health records, it is possible to define treatment discontinuations using pharmacoepidemiology methods. All prescriptions dispensed to these patients during the pre–follow-up period were identified, and the coverage of each prescription was calculated by dividing the total amount of drug filled in the prescription by the Defined Daily Dose \cite{WHOCollaboratingCentreforDrugStatisticsMethodology2003IntroductionResearch}. Subjects were defined as having discontinued treatment with MRAs during the first year of observation if they stopped filling the prescription and a minimum of 90 days had passed from the last day covered by the drug and the start of follow-up. The \texttt{R} package \texttt{adhereR}\cite{Dima2017ComputationData} was used to derive the discontinuation measure.  26 \%  of subjects resulted in having discontinued MRAs in the first year of therapy.

\subsection{Controlling for confounding}

The propensity score was estimated through logistic regression using variables that were possibly unbalanced in the two exposure groups. Matching was used to obtain a balanced dataset with respect to such variables. Descriptive statistics on the original and matched dataset are reported in Table \ref{matchingtable}. The matched cohort included 4598 patients. To test for the presence of unbalance, either the Chi-squared, Fisher or t-test were used as appropriate. Interestingly, even before matching, the two groups were already quite balanced. 

\begin{table}[!h]
\caption{Descriptive Statistics before and after matching.}
\centering
\begin{tabular}{lcccc}
  \hline
   & \multicolumn{2}{c}{\textbf{Before Matching}} & \multicolumn{2}{c}{\textbf{After Matching}} \\
 & \textbf{Stand. Effect Size}  & \textbf{p-value}  & \textbf{Stand. Effect Size}  & \textbf{p-value} \\ 
    \midrule
Sex & -2.90 & 0.62 & -0.35 & 0.92 \\ 
  Duration of HF & 0.41 & 0.94 & 4.92 & 0.17 \\ 
  BMI & -10.65 & 0.08 & -3.40 & 0.32 \\ 
  Blood Pressure & 9.74 & 0.10 & -3.21 & 0.37 \\ 
  High Heart Rate & -6.21 & 0.29 & 5.06 & 0.16 \\ 
  CPT & 15.54 & 0.01 & 2.31 & 0.44 \\ 
  Hypertension & 8.00 & 0.19 & -3.49 & 0.32 \\ 
  Anemia & 9.03 & 0.13 & -3.61 & 0.30 \\ 
  Vasculopathy & -3.01 & 0.61 & -0.28 & 0.94 \\ 
  BCPO & -1.09 & 0.85 & 0.40 & 0.91 \\ 
  Cancer & 0.53 & 0.93 & -0.11 & 0.98 \\ 
  Ulcer & 6.27 & 0.27 & -0.18 & 0.96 \\ 
  Liver disease & 0.27 & 0.96 & -3.61 & 0.34 \\ 
  Cerebrovascular diseases & 3.99 & 0.50 & -0.54 & 0.88 \\ 
  Dementia & 4.84 & 0.39 & -0.27 & 0.94 \\ 
  Mental Disorders & -1.75 & 0.77 & -4.90 & 0.23 \\ 
  Number of comorbidities $>$3 & 5.88 & 0.31 & -3.56 & 0.32 \\ 
  ACE/ARBs & -0.37 & 0.95 & -3.13 & 0.38 \\ 
  Beta-blockers & -6.01 & 0.31 & -2.64 & 0.47 \\ 
  Diuretics & 0.46 & 0.94 & 0.09 & 0.98 \\ 
  Digitalis & -3.37 & 0.57 & 5.17 & 0.13 \\
   \hline
\end{tabular}
\label{matchingtable}
\end{table}

\subsection{Results}

\begin{figure}[!ht]
    \centering
    \includegraphics[width=15cm]{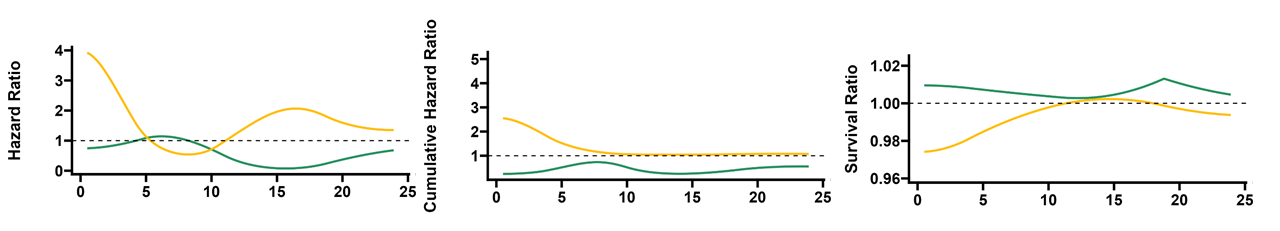}
    \caption{Final clusters medoids for the two-phenotype case. }
    \label{fig:appl1}
\end{figure}

The CATEs over a 2-year time frame according to the three effect measures (the hazard ratio, the cumulative hazard ratio and the survival ratio) were estimated using the SNnet-S. 100 bootstrap samples were obtained for each subject profile, and a maximum of four clusters was considered.
The average silhouette values and the internal agreement among bootstrap samples favoured two as the number of treatment effect phenotypes. This was consistent across the three measures of effect considered.
The average silhouette values and the internal agreement among bootstrap samples favoured two as the number of treatment effect phenotypes. This was consistent across the three measures of effect considered. For this reason, in Figure \ref{fig:appl1} the medoids of the final clusters for each effect measure for the two-phenotype case are displayed. For both the cumulative hazard ratio and the survival ratio, we see that there is a group (green line) that tends to show a stable protective effect and another that shows a higher risk of events in the first six months. In fact, in the latter group, this may also be reflected in the hazard, where there is an immediate adverse effect of discontinuation, which tends to decrease rapidly in the first half of the year. However, the overall agreement between the results of the different clusters for all three measures of effect is around 0.24, which means that, due to the sample size, the results are not robust enough to draw any firm conclusions.

\begin{figure}[!ht]
    \centering
    \includegraphics[width=15cm]{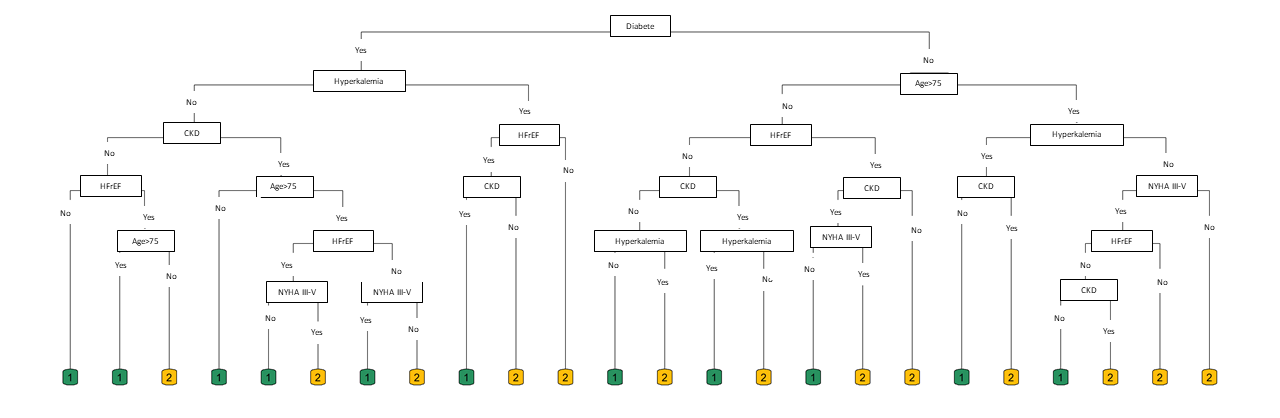}
    \caption{Subjects' assignment to treatment effect phenotypes, considering two as a number of clusters.}
    \label{fig:appl2}
\end{figure}

Nevertheless, as an example, the assignment of the subjects to the phenotypes for the survival ratio and the 2-clusters case is reported in Figure \ref{fig:appl2}.

\section{Discussion and Conclusions}

In this research, we have proposed a novel method to characterize which are the different responses to a treatment in a survival setting, taking into account two key aspects: effect heterogeneity and time-varying effects. Most of the methods to inform personalized treatment decisions are built on the assumption that treatment is either “beneficial”, “null”  or “harmful”. However, when considering a time interval as a decision frame of interest, there is a broader range of possibilities. Making decisions without having the full picture of the dynamic response of subjects can lead to unwanted consequences. On the other hand, among the infinite possibilities of responses that can be present in a population, it is essential to summarize to some extent which are the most common treatment-effect phenotypes over time and to which subject profiles they can be attributed. \\
The aim of the proposed approach is to tackle this issue by a two-stage procedure. 
First, the conditional average treatment effect over a time span of interest considering relevant effect modifiers was estimated using methods that allow both interactions and non-proportional hazards. Although a spline-based survival model can be useful when there is a strong prior knowledge of the model structure, the approach based on a survival neural network and smoothing of the subject-specific hazard curves is a valid alternative, as shown in the simulation study. \\
As a second step, an appropriate synthesis that encapsulates the possible responses to the treatment was obtained by clustering the curves representing the conditional treatment effects through functional clustering. While looking at the individual CATEs can be complicated to get a complete picture of the treatment behaviour in the population, clustering the subject-specific treatment effect curves can help to extract a useful summary. \\
Survival clustering has been already proposed in the literature to discover subpopulations whose survival is regulated by different generative mechanisms \cite{manduchi2022deep}. On the other hand, here we are clustering subject-specific treatment effects over time to discover subpopulations that display different treatment effect mechanisms.
Nevertheless, in future research, the method could be investigated with the aim of characterizing different risk patterns over time.   \\
We employed the term functional survival analysis to remark that we are treating the CATEs as realizations of functional objects.  To our knowledge, this aspect together with the idea of extracting treatment effect phenotypes over time has not been considered before in the literature. \\
Bootstrapping and consensus clustering were used to take into account the uncertainty in the two-stage procedures that involve the estimation and the clustering. An important aspect concerning the uncertainty that needs to be taken into account is that not all the subject profiles have the same prevalence in the population. Care should be taken in considering the subset of subject profiles for which data is available. In our simulation setting that considered five effect modifiers, the sample size had a strong effect on the accuracy of the clustering results. Indeed, a large number of subjects was needed in order to obtain results in line with the data-generating model. Specific sample size requirements are complex to obtain as they depend on the underlying model. Nevertheless, assessing internal agreement among the results on the different bootstrap samples can serve as a  quantification of clustering uncertainty. \\
In regard to the interpretation of the results, there are two different aspects to consider. The first regards the interpretation of the different phenotypes obtained with respect to the treatment effect over time. This can be done by visualizing the medoids of the different clusters. The second relevant component is which subjects are contained in each of the phenotypes. The mapping between the effect-modifiers space and the phenotypes can be represented by a decision tree, which can be easily used by domain experts to subjects' assignment to a specific phenotype. However, the interpretation of the effect of specific covariates on the CATEs is outside the scope of this work.
We have shown a possible application of the proposed approach on data coming from electronic health records and administrative health databases, involving the treatment of Heart Failure patients. In particular, the aim of the study was to assess the response to discontinuation of MRAs during the first year of treatment in the two consequent years. While this issue is very relevant for the treatment of heart failure,  the results reported in this work are intended only as exploratory. This application is meant only as an illustrative example of how the proposed method can be used to answer a medical decision question. Analyses on a  larger cohort are needed in order to confirm the results so that they can be of use in the clinical management of heart failure patients. An additional challenge in this context concerns the causal effect of interest. We have considered only an intention-to-discontinue effect, as subjects may have re-started treatment with MRAs during the consequent two years. Further directions of this analysis include also studying the per-protocol effect.  \\
With respect to the choice of the measure of effect, in survival analysis, Hazard Ratios are traditionally used. However, we have considered the ratio between the cumulative hazards and survival functions, as in the presence, of non-proportional hazards, they allow us to take into account the “accumulation” of risk over time. Although it can be useful to consider both of them, in the medical domain the ratio between the survival curves offers a clearer interpretation. Furthermore, depending on the specific problem, an absolute measure such as the survival difference could be employed instead.\\
A limitation is the need to choose a number of treatment-effect phenotypes. Even though there are established methods coming from the clustering literature, in this specific context it is also important to confirm the interpretability of the results from a domain-specific point of view. Moreover, the method relies on the choice of a specific clustering algorithm. Although we have considered functional k-mean, it would be possible to use any functional clustering algorithm. Indeed, consensus clustering is independent of the specific clustering method used, and it could in theory also be used to combine the results obtained from different algorithms without the need to choose one algorithm over the other. \\

\bibliographystyle{unsrt}  
\bibliography{references} 

\end{document}